\documentclass[runningheads]{svmult}

\usepackage{makeidx}   
\usepackage{graphicx}  
\usepackage{subeqnar}  
\usepackage{multicol}  
\usepackage{physprbb}  
\makeindex             

\begin{document}
\title*{A search for close companions in Sco~OB2}
\toctitle{A search for close companions in Sco~OB2}
%
%
\titlerunning{A search for close companions in Sco~OB2}
%
\author{Thijs Kouwenhoven\inst{1}
\and Anthony Brown\inst{2}
\and Hans Zinnecker\inst{3}
\and Lex Kaper\inst{1}
\and Simon Portegies~Zwart\inst{1,4}
\and Alessia Gualandris\inst{1,4}}
\authorrunning{Thijs Kouwenhoven et al.}
%
%
\institute{Astronomical Institute ``Anton Pannekoek'', 
  University of Amsterdam, Kruislaan 403, 1098 SJ, Amsterdam, The Netherlands
\and Leiden Observatory, University of Leiden, 
     P.O. Box 9513, 2300 RA Leiden, The Netherlands,
\and Astrophysikalisches Insitut Potsdam, An der Sternwarte 16, 
     D-14482, Potsdam, Germany,
\and Section Computational Science, University of Amsterdam, 
     Kruislaan 403, 1098 SJ, Amsterdam, The Netherlands}

\maketitle              

\begin{abstract}
  Using adaptive optics we study the binary population 
  in the nearby OB~association Scorpius~OB2. 
  We present the first results of our near-infrared adaptive 
  optics survey among 199 (mainly) A- and B-type stars 
  in Sco~OB2.
  In total 151 components other than the target stars 
  are found, out of which 77 are probably background stars.
  Our findings are compared with data collected from 
  literature. Out of the remaining 74 candidate physical 
  companions 42 are new, demonstrating that many stars
  A/B stars have faint, close companions.
\end{abstract}

\section{The primordial binary population in Sco~OB2}
\begin{figure}[bt]
  \begin{minipage}[hbt]{\textwidth}
    \hbox{\hspace{0cm}
    \includegraphics[width=0.56\textwidth,height=!]{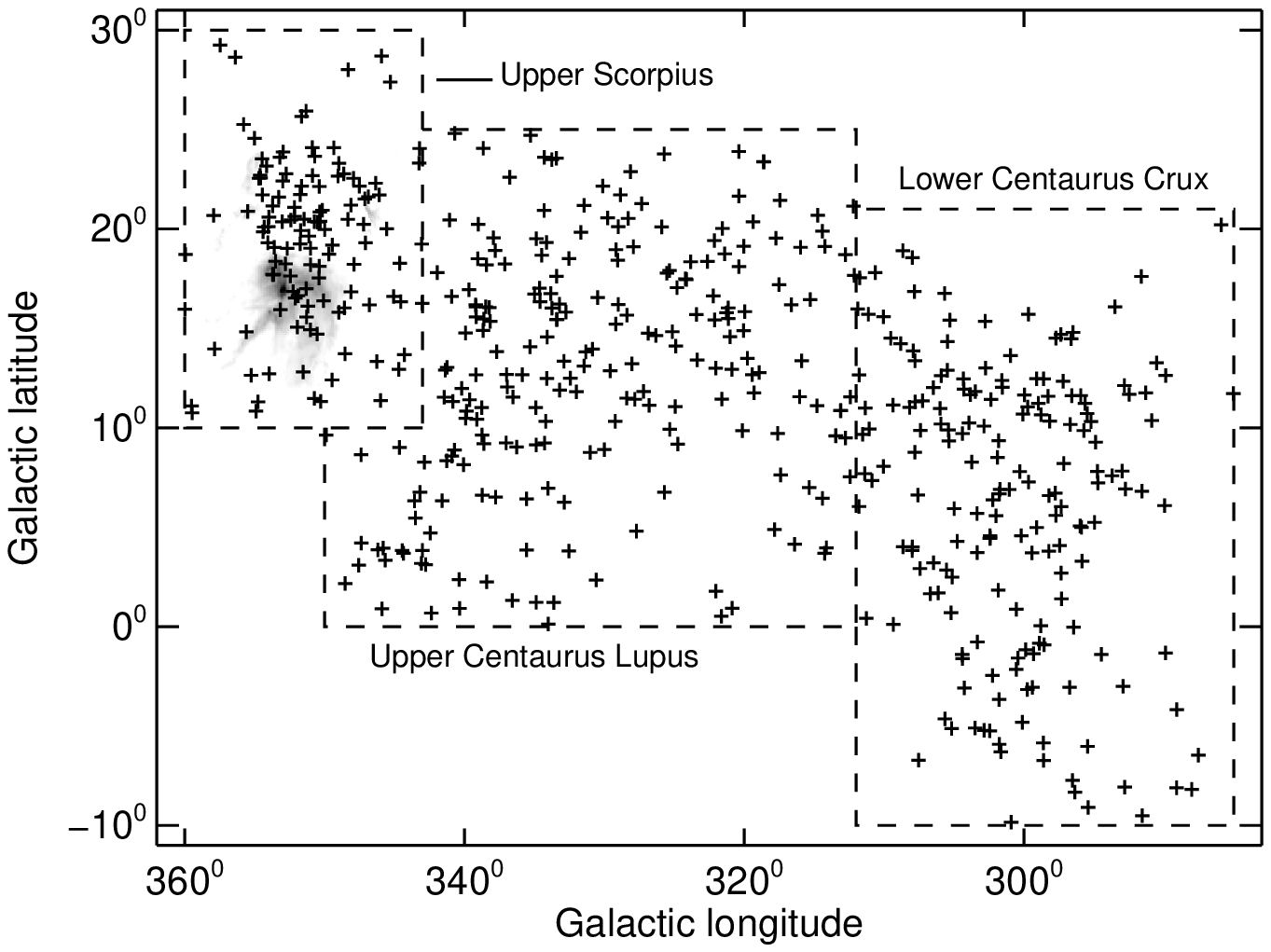}\hspace{0cm}
    \includegraphics[width=0.44\textwidth,height=!]{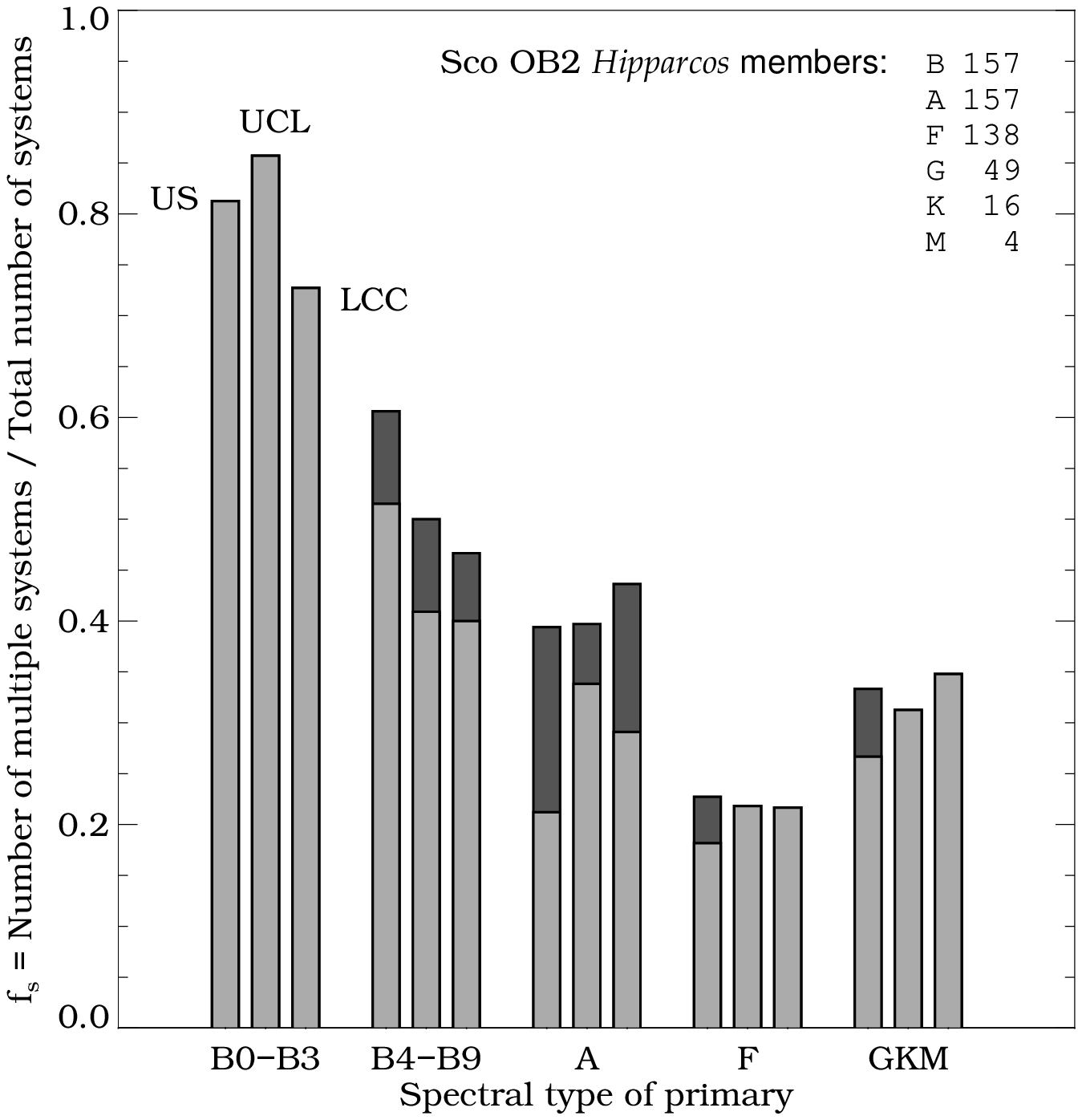} }
    \caption[]{ {\it Left:} The three subgroups of Sco~OB2. The $\rho$~Oph star forming 
region is visible in the upper-left corner in grey scale (IRAS 100~$\mu$m). {\it Right:} The fraction of stellar systems which is multiple versus the spectral type of the primary, for the three subgroups of Sco~OB2. Only confirmed Hipparcos member primaries are considered here. The light and dark grey parts of the bars correspond to literature data and the new data presented in this article, respectively.  \label{figure: eps}}     
    \hfill
  \end{minipage}
\end{figure}

The primordial binary population (PBP) is defined as the population of binaries as established just after the gas has been removed from the forming system, i.e. when the stars can no longer accrete gas from their surroundings \cite{kouwenhoven2003}. 
Characterizing the PBP is important for our understanding of the process of star formation, the formation and evolution of OB associations, the origin of the field star population and OB runaway stars, and the production and evolution of binary systems.
OB associations are ideal sites for the study of the 
PBP. Since OB associations are young ($5-50$~Myr) and low density ($\approx 0.1~{\rm M}_\odot~{\rm pc}^{-3}$) stellar systems, the effects of stellar and dynamical evolution are modest. OB associations are practically cleared of gas and the full stellar population (from OB stars to brown dwarfs) is present.
Sco~OB2 (Figure~\ref{figure: eps}) is the closest young OB association, its proximity facilitating the detection of close and faint companions. The membership and stellar content has been established with {\it Hipparcos} \cite{dezeeuw1999}, and many binary surveys have been performed in the past.

\section{Adaptive optics observations}

We performed a near-infrared adaptive optics (AO) survey among 199 (mainly) A- and B-type stars. Our sample is a subset of the {\it Hipparcos} membership list of Sco~OB2 \cite{dezeeuw1999}. 
The AO observations bridge the observational gap between the wide visual and close spectroscopic binaries. In the near-infrared the luminosity contrast between the primary star and its (often later-type) companion(s) is lower, which facilitates the detection of faint and close companions.
The observations were performed with the ADONIS/SHARPII+ system. This AO system was mounted on the ESO 3.6m telescope on La Silla. The camera field of view is $12.8 \times 12.8$~arcsec. Each star is observed at four different pointings in order to maximize the available field of view. The angular separation between the target stars and the other components in the field ranges between 0.2~and 15~arcsec. All target stars are observed in the $K_S$ band, for some also $J$ and $H$ band observations are obtained. 
We applied standard data reduction techniques in combination with image selection, retaining only the highest Strehl ratio images. We find a total of 151 stellar components in the fields around the 199 target stars. For each detected component the projected distance, position angle, and relative magnitude are measured.

The status of an observed component (companion star or background star) can be determined using the colour-magnitude diagram: companion stars are expected to lie on the isochrone, while background stars are not \cite{shatsky2002}. For most of our target stars only the $K_S$ magnitude is available to determine whether a component is a companion star or a background star. We have used a simple brightness criterion to separate background and foreground stars. We consider all stars fainter than $K_S=12$ to be background stars, and the stars brighter than $K_S=12$ companion stars (the magnitude of an M5V star at the distance of of Sco~OB2 is $K_S=12$). Using this method we find 74 candidate companion stars and 77 probable background stars.

\section{Binary statistics and observational biases}

We combine our observations with literature data on binary in Sco~OB2. This dataset includes visual, astrometric, spectroscopic, and eclipsing binaries. A careful comparison shows that 42 out of the 74 candidate companion stars that we found are new: 14 in US, 14 in UCL and 14 in LCC (Table~\ref{table: statistics}).
It was pointed out by \cite{brown2001} that the observed number of multiple systems with A- and late B-type primaries is relatively low due to observational biases. Part of this selection effect is now removed with our new AO observations (Figure~\ref{figure: eps}).
\begin{table}[bt]
  \centering
  \setlength{\tabcolsep}{0.7\tabcolsep}
  \begin{tabular}{l cc cccc cc c}
    \hline
    Subgroup              & $D$ (pc) & Age (Myr) & $N_s$ & $N_b$ & $N_t$ & $N_m$ & $f_s$ & $f_c$ & $CSF$ \\
    \hline
    Upper Scorpius        & 145  & 5    & 63    & 46    & 7     & 3     & 0.47  & 0.67  & 1.61\\
    Upper Centaurus Lupus & 140  & 13   & 131   & 68    & 17    & 4     & 0.40  & 0.61  & 1.52\\
    Lower Centaurus Crux  & 118  & 10   & 112   & 54    & 13    & 0     & 0.37  & 0.57  & 1.45\\
    \hline
    Scorpius OB2          &      &      & 303   & 171   & 37    & 7     & 0.41  & 0.61  & 1.52\\
    \hline
  \end{tabular}
  \caption{Multiplicity among {\it Hipparcos} members of Sco~OB2. The columns show the subgroup names, their distances \cite{dezeeuw1999}, the ages \cite{degeus1989}, the number of known single stars, binary stars, triple systems and $N>3$ systems. The last three columns show the binary statistics: the fraction of multiple systems $f_{\rm s} = (N_b+N_t\dots) \ /\ (N_s+N_b+N_t+\dots)$, the fraction of stars in multiple systems $f_{\rm c} = (2N_b+3N_t\dots) \ /\ (N_s+2N_b+3N_t+\dots)$ and the companion star fraction $CSF = (2N_b+3N_t\dots) \ /\ (N_s+N_b+N_t+\dots)$, which measures the average number of companion stars per primary star. \label{table: statistics} } 
\end{table}

\section{Future work}

The ultimate goal of our study is to determine the PBP. We will achieve this by combining the binary data of Sco~OB2 with detailed numerical models. These models are N-body simulations including state of the art stellar and binary evolution \cite{spz2001}. Simulated observations will be produced to characterize in detail the selection effects for the different binary surveys. These simulations will also be used to investigate the impact of dynamical and evolutionary effects that may have altered the binary population over the lifetime of Sco~OB2. 
Combined with the knowledge about the selection effects the PBP can then be reconstructed. The results will provide the characteristics of the binary population as a function of environment just after the star formation process has finished. This is an important constraint on theories of star (cluster) formation.

%

\end{document}